\documentclass{mem}
\usepackage{natbib}\usepackage{txfonts}\usepackage{balance}
\usepackage{graphicx}
\idline{75}{282}
\begin{document}
\def\teff{$T\rm_{eff }$}
\def\kms{$\mathrm {km s}^{-1}$}

\title{
The mystery of the telescopes in Jan Brueghel the Elder's paintings
}

   \subtitle{}

\author{
P. \,Molaro\inst{1} 
\and P. \,Selvelli\inst{1}
          }

  \offprints{P. Molaro}

\institute{
Istituto Nazionale di Astrofisica --
Osservatorio Astronomico di Trieste, Via Tiepolo 11,
I-34143 Trieste, Italy.
\email{molaro@oats.inaf.it}
  }

\authorrunning{Paolo Molaro }

\titlerunning{The  Jan Brueghel the Elder's spyglasses}

\abstract{
Several  early spyglasses 
 are depicted  in   five paintings  by    Jan Brueghel the Elder
completed between 1608 and 1625,  as he was  court painter of   Archduke Albert VII of Habsburg.
An optical    tube  that appears  in the  {\it Extensive Landscape with View of the Castle of Mariemont},  dated 1608-1612,  represents  the first painting of a telescope whatsoever. We collected some  documents 
 showing  that Albert VII  obtained   
spyglasses  very early  directly from   Lipperhey or Sacharias Janssen.  Thus the  painting likely
  reproduces one of the first man-made telescopes ever. Two   other instruments 
 appear  in  two   Allegories of Sight made in the years 1617 and 1618. These are sophisticated  instruments and the  structure   suggests that they may  be  keplerian, but this is about two decades  ahead  this mounting  was in use. 
\keywords{Invention of telescope, keplerian telescopes, Jan Brueghel, Galileo Galilei, International Year of Astronomy 2009}
}
\maketitle{}

\section{Introduction}

Spyglasses and  other astronomical
instruments  are present in   five  paintings, one landscape and four  allegories,  that  Jan Brueghel the Elder
(1568-1625) painted in the years  between 1608 and 1625, often in collaboration  with  P. P. Rubens. 
The paintings were executed  while the artist was  court painter  of  Archduke Albert VII of Habsburg (1559-1621), the Spanish
Governor of the catholic part of Netherlands,  who had  genuine scientific interests.   All  paintings    are   high quality, detailed and realistic  pictures    of these instruments  offering interesting clues on the early evolution of the telescope where documents are scarce.   Previous studies  on this subject can be found   in \citet{selvelli97}, \citet{selvelli09} and \citet{molaro09}.

\section{The first painting of a  spyglass} 

A particular of the painting  {\it Extensive Landscape with View of the Castle of Mariemont }  conserved at the Virginia Museum of Fine Arts in Richmond, VA,
USA is shown in Fig.1.  The painting does not report a date  but from    the development of the works    of the Mariemont Castle   the painting    has been dated   between 1608 - 1611.
In the detail shown in the figure   
the Archduke Albert VII is watching the landscape  through a spyglass.
The instrument has a cylindrical shape and appears to be   metallic   with two  gilded rings   on both sides.   The  length is  about 40-46 cm  and the  diameter of about 5 cm. 
To our knowledge,  this painting  represents the most ancient reproduction
of a  spyglass. 

We provide here  several documents  which  link this early instrument with one made by the still unknown inventor of the telescope.
One first evidence comes from  Guido Bentivoglio, the Papal nuncio at the court of Albert VII,   who was quite close to the devote  Archduke
and  was present when Spinola, Commander of the Spanish Army in the
Flanders, came back from The Hague after having witnessed   the first public demonstration of a telescope on  25th September  1608.
Spinola was in The Hague in that
period, as  representative  of the Spanish Governor  for peace negotiations
with the {\it Staatsholder }  of the seven provinces prince Maurice of Nassau. The Truce  was actually signed in April 1609.
 On the  2nd of  April 1609  Bentivoglio in a letter  to Cardinal Scipione Borghese, the  nephew of Paul V and papal secretary,  wrote:
 {\it When the marquis Spinola returned from HollandÉÉ.. the Archduke and the Marquis himself were most desirous to obtain such an instrument, and indeed it came about that one came into their hands, although not of such perfection as the one owned by Count Maurice } (cfr \citet{hensen23} for the entire text). 
A second evidence comes from Daniello Antonini,  a noble from Udine and friend of Galileo, who   was serving in the   Archduke army
in Brussels. In
September 1611 Antonini   wrote a letter to Galileo
telling  him that the Archduke owned  some spyglasses  obtained from the inventor:  {\it Ho veduti de' piu' esquisiti occhiali che si fabrichino in queste 
parti. NÕ ho veduti di quegli del proprio primo inventore, dati poi a questo 
Serenissimo, ma son tutti dozinali}.  Also  
Maria Schyrlaeus de Rheita in  its {\it  Oculus Enoch et
Eliae}   (1646, p. 337)  wrote that
the Marques Ambrogio Spinola bought a spyglass  in The
Hague near the end of 1608, probably made by Lipperhey,  and
offered it  to  Archduke Albert.  
 %The manufacturer could be either   Lipperhey or Janssen, both considered as possible {\it fathers} of the telescope.
On the other hand  Pierre Borel  in the  De Vero Telescopii Inventore (1656)  
 quotes    the son of Sacharias Janssen's   declaration, made in 1655   to the Middelburg City Council in an investigation about  the  origin of the telescope suggesting that the manufacturer is Janssen. The declaration sounds:
{\it  Our artisan [Sacharias Janssen] first made tubes of 16 inches, and gave the best to Prince Maurice and Archduke Albert, as we shall see below in the testimonies, for which he received money and was asked not to divulge the thing further.} 
Thus this documentation   shows that it is very likely that  the  optical tube held by the archduke in the {\it Extensive Landscape with View of the Castle of Mariemont } represents one
of the  spyglasses belonging to Albert VII   obtained  directly from the inventor of the telescope.

%We note that the presence of the spyglass in the painting can be used to exclude the
%year 1608  in the range of the dates for its completion, since the first
%spyglasses  appeared in autumn 1608, while  the landscape with the trees
%and other details   in the painting  indicates  a late summer as the more likely
%season. 

 \begin{figure}[]
%\begin{figure*}[t!]
\resizebox{\hsize}{!}{\includegraphics[clip=true]{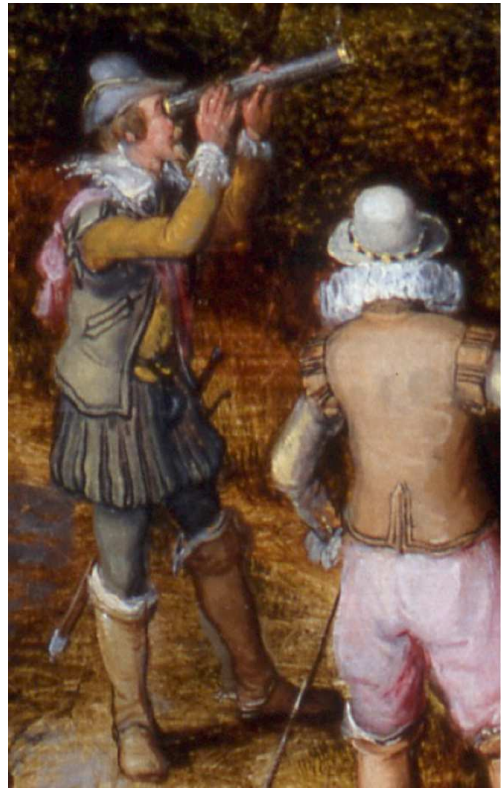}}
\caption{\footnotesize
 Detail of the  "Extensive Landscape with View of the
Castle of Mariemont" by J. Brueghel the Elder, ca. 1608-1612.  
Museum of Fine Arts, Richmond, Virginia.
%The Adolph D. and Wilkins C. Williams 
%Fund. Photo: Ron Jennings. 
}
\label{fig1}
%\end{figure*}
\end{figure}

\section{The silver telescopes}

{\it The Allegory of Sight} is one of the five  paintings of  the  series of the {\it Allegory
of the Senses},   made   in
collaboration with  Peter Paul Rubens,   which  can be admired at 
the Museum El Prado in Madrid.
% (Belloni  1964).
The painting, oil on wood,  depicts a hall
in the ancient royal Palace of
Brussels,  residence of the Archdukes, on the hill of Coudemberg,
where  paintings, precious items, and
scientific instruments were collected. 
One can note    a
large astrolabe, an armillary sphere, a pedestal globe, a
pair of Galileian compasses, map dividers and sundials.
  Each instrument was  meticulously
characterized with true Flemish skill so that even the most minute details
are accurately reproduced.
The
painting had been  completed by  1617 as testified by the
date   on a roll of  papers lying over the  book 
{\it Cosmographie}  in the
lower part of the painting 
 besides the author's signature.  
The telescope  between Venus and Cupid shown in a detail in  Fig. 2,  consists
of a main
tube and seven draw-tubes  which appear to be made of metal.
 The  draw tubes terminates
in  enlarged collars   made of the same material and the lenses are housed in large rounded terminals. 
The instrument is fixed into a curved metal sleeve support attached to a
adjustable brass joint.  
%The pedestrial  consists of a
%turned column terminating in a simpler saucer-shaped round base
%(cfr Bedini 1971).
A comparison with other objects depicted in the painting  indicates   a
maximum width for the draw tubes of about 6-7 cm  and a minimum  of 2 cm in proximity of the eyepiece. The  total length with the tubes  all drawn and considering a tilt of about 30 degrees along the line of sight is of
about 170-180 cm.

A   similar telescope is reproduced in the   {\it The Allegory of the Sight and the Sense of
Smell}, an oil on
canvas of considerable size,  176 for 264 cm.  This  painting,  completed around the same period
(1618-1620), was commissioned by the City of
Antwerp to Jan Breughel   to celebrate the
visit of  the
Archdukes.   About 12 painters, including P.P. Rubens,  contributed  to the  painting inaugurating
 the 
{\it kunstkammer } style, that became  fashionable afterwards. However, the  painting exposed at the Prado    is  a copy of the
original
that
went lost in the fire of the Castle of Coudemberg  in 1731.   The painting includes several of the instruments
  reproduced in the other {\it Allegory of Sight } of 1617, but  the two telescopes differ in several important details. The number of
draw tubes is   eight  and not  seven,  and    the rings are colored.
The   similarity   
indicates    the same maker,  but they are likely
two different  instruments.

 \begin{figure}[]
%\begin{figure*}[t!]
\resizebox{\hsize}{!}{\includegraphics[clip=true]{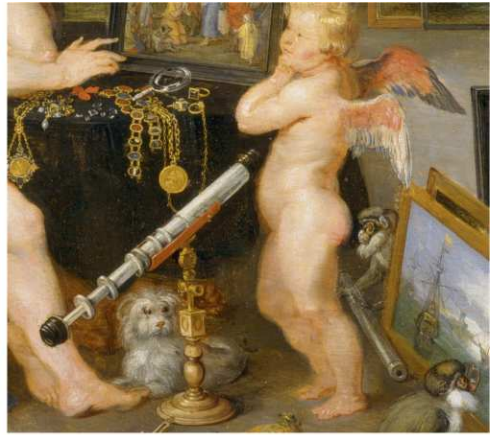}}
\caption{\footnotesize
  Detail of the  "The Allegory of Sight" by J. Bruegel and
P.P. Rubens,  1617.  Museo Nacional del Prado, Madrid. }
\label{fig2}
%\end{figure*}
\end{figure}

\subsection{First  keplerian telescopes?}

 The  technology  of these telescopes is quite advanced    for the epoch since there is  no record of similar instruments (cfr the Catalogue of Early
Telescopes  by \citet{vanhelden99}).  The closest resemblance 
is found with the illustrations reported in Christopher  Scheiner's works
({\it Disquisitiones Mathematicae} 1614, and  {\it Rosa Ursina}  of 1631).
More surprisingly,  the telescopes depicted  may represent   first 
examples of  keplerian telescopes.
 The origin and development of a  telescope 
consisting of two convex lenses is  uncertain and open to question.  It was theoretically  described by Kepler in his {\it  Dioptrice} 
in  
1611 but     Kepler did not make it and we have to wait till  C. Scheiner 's {\it Rosa Ursina }  (1631)   for the first
book containing  a description of   an astronomical telescope.
Francesco Fontana  in his   {\it Novae Celestium Terrestriumque rerum Observationis}  (1646) claimed  to have manufactured an astronomical 
telescope already in 1608. In support Fontana produced   a declaration of father Zupo  
stating  that,  together with father Staserio, he  saw  a  two convex lenses telescope made by him  in   1614.
 
Thus the existence of keplerian telescopes already around 1617-1618 is quite  remarkable.
Three circumstantial
considerations seem to support  this  idea.     
%When fully opened the overall length of the telescope   is estimated to be about 180 cm.
 Long telescopes about 2 m are much easier to make in the keplerian mounting and with a Dutch mounting  these long  telescopes   would imply an unpractically small   field of
view of few arcminutes. The telescopes show a small minimal width of about 2 cm   in the proximity of the eyepiece, which  is also 
 easier to be obtained with a keplerian mounting  than with a Doutch munting where the beam size never decreases below the size with it reaches  the concave length.
But the most important element is the  presence of  quite large eyepieces.   With a  negative lens   the eye needs 
to be  brought as close as possible to the lens since  the eye's pupil 
becomes the aperture stop and the exit pupil.  With a convex lens as eyepiece the eye has to be
positioned  to its focus and the structure of the eyepiece is manufactered just to help the eye positioning. 
Finally, the first records of Keplerian  telescopes are somewhat related to the Habsburgs  family.
Cristopher Scheiner in  his {\it Rosa Ursina } claimed  that he made a keplerian
instrument in 1614-1617 and showed it to  Archduke Maximilian III, 
brother of  Albert VII.  According to documents in the Tyrolean State Museum (cfr Daxecker 2004, p13,14),  around 1615, Maximilian 
 received a telescope with two
convex lenses and Scheiner added a third one,  thus manufacturing a
terrestrial keplerian telescope.  We note, incidentally, that Scheiner
actually  used a Dutch telescope for his observations of 
sunspots in 1611 and it is not clear when he started with  a keplerian one. it was certainly not  before 1615 as it appears from his manuscript {\it Tractatus de Tubo Optico} of 1615 (\citet{daxecker01}), but more likely only  after 1624   (\citet{vanhelden77}).
It is quite possible that Albert VII obtained from his brother a keplerian telescope     for his collection.

 %\subsection{The monkey's tube}
 
In the  painting in  Fig. 2, on the floor  just behind Cupid  one can note a tube held by
a monkey. A close inspection  shows  that  the tube  is also 
 a spyglass. It  is remarkable that it belongs to the Archduke's collection and it could be the same spyglass depicted  in Fig. 1 several years before.  
 The monkey holding the tube while another monkey is holding two glasses   has   certainly an allegorical meaning.
 In  Flemish painting the monkey  is a  traditional symbol for foolishness, and here it may underline the 
 brain-storming implications of the new discoveries or, more simply,  
 the serendipitous way in which  the   telescope was  conceived.

\begin{acknowledgements}
We gratefully thank Virginia
Museum of Fine Arts,   the Adolph D. and Wilkins C. Williams 
Fund., Howell  Perkins and Ron Jennings for the  reproduction  of the painting  and  Inge Keil and Franz Daxecker  for helpful information.
 \end{acknowledgements}

\bibliographystyle{aa}

\begin{thebibliography}{}

\bibitem[(2000)]{bedini}
%        Bedini, S.A.  1971,  Phisis,    Vol. 13, 149
 % \bibitem[(2000)]{belloni}
  %      Belloni, L. 1964, Rendiconti Istituto Lombardo, B98, 238
  \bibitem[Hensen (1923)]{hensen23} Hensen, A. H. L., "de Verrekijkers van Prins Maurits en van Aartshertog Albertus",  Mededeelingen van het nederlandisch Historisch Instituut te Rome,  (1923) 199
  \bibitem[Daxecker (2001)]{daxecker01} Daxecker, F.  2001 Beitrage zur Astronomiegeschichte  Bd. 4, S 19-32
  \bibitem[Daxecker (2004)]{daxecker04} Daxecker, F.  2004 "The Physicist and Astronomer Christopher Scheiner",  Innsbruck University  publ.  N. 246
  \bibitem[Molaro \& Selvelli (2009)]{molaro09} Molaro,P., Selvelli,P., 2009, The Role of Astronomy in Society and Culture IAU Symp  260, D. Valls-Gabaud \& A. Boksenberg, eds.
  \bibitem[Selvelli (1997)]{selvelli97}
        Selvelli, P.  1997 L' Astronomia, 175, 36
   \bibitem[Selvelli \& Molaro (2009)]{selvelli09} Selvelli, P., Molaro, P.,  2009  Proceedings 400 Years of Astronomical Telescopes:
A Review of History, Science and Technology, ESA/ESTEC
Noordwijk, The Netherlands     
\bibitem[Van Helden (1977)]{vanhelden77}
  Van Helden 1977,
         Transa. of the American Philosophical Society 67, no. 4,
 \bibitem[Van Helden (1999)]{vanhelden99}
        Van Helden, 1999,  Catalogue of early telescopes Giunti, Firenze, 
\end{thebibliography}

\end{document}